\documentclass[letterpaper, 10 pt, conference]{IEEEtran}
\IEEEoverridecommandlockouts                              
\usepackage{setspace}
\usepackage{pgfornament}
\usepackage{myStyle}
\usepackage{subcaption}

\usepackage{eso-pic}
\AddToShipoutPictureBG*{%
	\AtPageUpperLeft{%
		\setlength\unitlength{1in}%
		\hspace*{\dimexpr0.5\paperwidth\relax}
		\makebox(0,-0.5)[c]{\begin{tabular}{c c}
				Max van Haren \textit{et al}, 	A Frequency-Domain Approach for Enhanced Performance and Task Flexibility in Finite-Time ILC. \\
				To appear in 22nd European Control Conference (ECC), uploaded to arXiv \today \\
		\end{tabular}}
}}

\begin{document}
	\bstctlcite{BSTControl}
	\title{\fontsize{23.5}{28.2}\selectfont A Frequency-Domain Approach for Enhanced Performance and Task Flexibility in Finite-Time ILC}

\author{Max van Haren$^{1}$, Kentaro Tsurumoto$^{2}$, Masahiro Mae$^{2}$, \\
	Lennart Blanken$^{1,3}$, Wataru Ohnishi$^{2}$ and Tom Oomen$^{1,4}$
	\thanks{This work is part of the research programme VIDI with project number 15698, which is (partly) financed by the Netherlands Organisation for Scientific Research (NWO). In addition, this research has received funding from the ECSEL Joint Undertaking under grant agreement 101007311 (IMOCO4.E). The Joint Undertaking receives support from the European Union Horizon 2020 research and innovation programme. Furthermore, it is partially funded by the joint JSPS-NWO funding programme \textit{Research Network on Learning in Machines}.}
	\thanks{$^{1}$Max van Haren, Lennart Blanken and Tom Oomen are with the Control Systems Technology Section, Department of Mechanical Engineering, Eindhoven University of Technology, Eindhoven, The Netherlands, e-mail: {\tt\small m.j.v.haren@tue.nl}.}%
	\thanks{$^{2}$ Kentaro Tsurumoto, Masahiro Mae and Wataru Ohnishi are with the department of Electrical Engineering and Information Systems, the University of Tokyo, Tokyo, Japan.}
	\thanks{$^{3}$ Lennart Blanken is with Sioux Technologies, Eindhoven, The Netherlands.}
	\thanks{$^{4}$Tom Oomen is with the Delft Center for Systems and Control, Delft University of Technology, Delft, The Netherlands.}}	
	\maketitle
	 \thispagestyle{empty}
	\begin{abstract}
		Iterative learning control (ILC) is capable of improving the tracking performance of repetitive control systems by utilizing data from past iterations. The aim of this paper is to achieve both task flexibility, which is often achieved by ILC with basis functions, and the performance of frequency-domain ILC, with an intuitive design procedure. The cost function of norm-optimal ILC is determined that recovers frequency-domain ILC, and consequently, the feedforward signal is parameterized in terms of basis functions and frequency-domain ILC. The resulting method has the performance and design procedure of frequency-domain ILC and the task flexibility of basis functions ILC, and are complimentary to each other. Validation on a benchmark example confirms the capabilities of the framework.
	\end{abstract}

\section{Introduction}
The increasing requirements for precision mechatronics result in a situation where both tracking performance and task flexibility, which is the ability to be flexible for different references, are important. Feedforward control is effective in compensating known disturbances for systems, leading to improved performance. Feedforward control is often based on models \cite{Butterworth2012}, which is generally achieved in industrial applications by means of basis functions feedforward, where the feedforward signal is a linear combination of basis functions that relate to physical quantities, such as acceleration feedforward for the inertia \cite{Boerlage2004,Lambrechts2005,Oomen2019}. Due to modeling and tuning inaccuracies, the increasing requirements for performance are generally not achieved.

Iterative Learning Control (ILC) can improve tracking performance with respect to model based feedforward control \cite{Bristow2006}, and hence, can fulfill the increasing requirements for performance. ILC utilizes information from past iterations to improve the tracking performance in the current iteration. For ILC to be industrially applicable, it is required that ILC
\begin{enumerate}
	\item[(R1)] is task flexible;
	\item[(R2)] has high tracking performance; and
	\item[(R3)] has an intuitive design procedure.
\end{enumerate}
In this paper, two types of ILC are considered, and are referred to as frequency-domain and norm-optimal ILC.

First, frequency-domain ILC uses infinite-time frequency-domain system representations to iteratively update the feedforward \cite{Arimoto1984}. Frequency-domain ILC has the advantage that convergence can be verified and tuned using frequency response function (FRF), that are accurate and inexpensive to obtain \cite{Pintelon2012}, leading to an intuitive design procedure and high tracking performance. Frequency-domain ILC is typically implemented in finite-time, where convergence can still be analyzed \cite{Norrlof2002}. However, conventional frequency-domain ILC is not directly capable of task flexibility, and hence, does not satisfy requirement R1.

Second, norm-optimal ILC utilizes a finite-time cost function to iteratively optimize the feedforward signal \cite{Gunnarsson2001}. The main advantage of norm-optimal ILC is that the feedforward signal can be parameterized into basis functions, that enables task flexibility \cite{Phan1996,VanDeWijdeven2010}. However, if the basis functions are not sufficiently rich to describe the inverse system, the performance is significantly worse compared to frequency-domain ILC, and therefore does not achieve requirement R2.

Important developments have been made to combine the task flexibility of basis functions ILC with the performance of frequency-domain ILC. In \cite{Mishra2009,Boeren2016}, frequency-domain ILC is projected on basis functions, resulting in task flexibility, but reducing tracking performance. Third, in \cite{Tsurumoto2023}, frequency-domain ILC is combined with basis functions ILC using a sequential optimization problem, that results in the performance of frequency-domain ILC and the task flexibility of basis functions ILC, but resulting in an unintuitive design procedure, not satisfying requirement R3. 

Although ILC methods with high performance and task flexibility are investigated, a method that achieves both high performance and task flexibility, with an intuitive design procedure, is currently lacking. In this paper, high performance (R2) is achieved by deliberately overparameterizing the feedforward in a low number of basis functions to accomplish task flexibility (R1) and a complementary signal that is optimized via an intuitive frequency-domain ILC design (R3). The key contributions in this paper include the following.
\begin{itemize}
	\item[(C1)] Determining the equivalent norm-optimal finite-time representation of frequency-domain ILC by specific choice of weighting matrices, that enables intuitive tuning in the frequency-domain for norm-optimal ILC (R3) (\secRef{sec:fILCNO}).
	\item[(C2)] Achieving both task flexibility (R1) and high performance (R2) by exploiting an overparameterized feedforward signal, building upon the norm-optimal description of frequency-domain ILC from C1 (\secRef{sec:fILCBF}).
	\item[(C3)] Validation of the framework on an example (\secRef{sec:example}).
\end{itemize}
\paragraph*{Notation} 
Let $\mathcal{H}(z)$ denote a discrete-time, Linear Time-Invariant (LTI), single-input, single-output system. The frequency response function of $\mathcal{H}(z)$ is obtained by substituting $z=e^{j\omega} \:\forall \omega\in [0,2\pi)$, and is denoted by $\mathcal{H}(e^{j\omega})$.

Signals are of length $N$. Vectors are denoted as lowercase $x$ and matrices as uppercase $X$. The $z$-transform of signal $x(k)$ is $\mathcal{X}(z)=\sum_{k=0}^{\infty}x(k)z^{-1}$. Let $h(k) \: \forall k\in\mathbb{Z}$ be the impulse response coefficients of $\mathcal{H}(z)$, with infinite impulse response $y(k) = \sum_{\tau=-\infty}^{\infty}h(\tau)u(k-\tau)$. Let $u(k)=0$ for $k<0$ and $k\geq N$ to obtain the finite-time convolution
\[
\resizebox{\hsize}{!}{$
	\displaystyle
\underbrace{\begin{bmatrix}
	y(0) \\ y(1) \\ \vdots \\ y(N-1)
\end{bmatrix}}_y = \underbrace{\begin{bmatrix}
	h(0) & h(-1) & \cdots & h(-N) \\
	h(1) & h(0) & \cdots & h(-N+1) \\
	\vdots & \vdots & \ddots & h(-1) \\
	h(N) & h(N-1) & \cdots & h(0) \\
\end{bmatrix}}_H\underbrace{\begin{bmatrix}
u(0) \\ u(1) \\ \vdots \\ u(N-1)
\end{bmatrix}}_u,$}
\]
with $y,u\in\mathbb{R}^N$ and $H\in\mathbb{R}^{N\times N}$ is the finite-time convolution matrix corresponding to $\mathcal{H}(z)$
 
\section{Problem Formulation}
\label{sec:pdef}
In this section, the problem that is dealt with in this paper is formulated. First, the problem setup is presented. Second, the different classes of ILC considered in this paper are described. Finally, the problem that is addressed in this paper is defined.
\subsection{Problem Setup}
\begin{figure}[tb]
	\centering\includegraphics{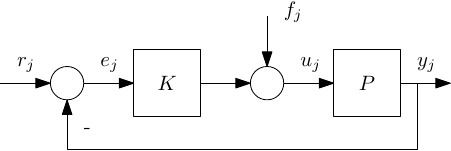}
	\caption{Control structure considered.}
	\label{fig:controlStructure}
\end{figure}

The control structure is seen in \figRef{fig:controlStructure}.
The LTI system $P$ is stabilized by LTI feedback controller $K$. The finite-time reference $r_j\in\mathbb{R}^N$ can be trial varying. The goal is to reduce the reference induced error $e_j\in\mathbb{R}^N$ over multiple trials $j$ with the trial-varying feedforward signal $f_j\in\mathbb{R}^N$.
\subsection{Classes of ILC}
In this section, the three considered classes of ILC, that is norm-optimal ILC, basis functions ILC and frequency-domain ILC, are presented.
\subsubsection{Norm-Optimal ILC}
Norm-optimal ILC is a type of ILC that minimizes a finite-time cost function, typically
\begin{equation}
	\label{eq:NOILCCost}
	\min_{f_{j+1}^{NO}} \left\|\hat{e}_{j+1}\right\|_{W_e}^2 \!+\left\|f_{j+1}^{NO}\right\|_{W_f}^2 \!+\left\|f_{j+1}^{NO}-f_j^{NO}\right\|_{W_{\Delta f}}^2,
\end{equation}
where is used that $\|x\|_W^2 = x^\top W x$, $W_e$, $W_f$ and $W_{\Delta f}$ are positive (semi)definite weighting matrices \cite{Gunnarsson2001}, $\hat{e}_{j+1} = e_j-\hat{J}\left(f_{j+1}-f_j\right)$, and finite-time convolution matrix $\hat{J} = \hat{P}\left(I+K\hat{P}\right)^{-1}$, with model $\hat{P}\in\mathbb{R}^{N\times N}$ and identity matrix $I\in\mathbb{R}^{N\times N}$. The optimal solution to \eqref{eq:NOILCCost} is of the form
\begin{equation}
	\begin{aligned}
		f_{j+1} = Q^{NO}f_j+L^{NO}e_j,
	\end{aligned}
\end{equation}
with norm-optimal ILC robustness and learning matrices $Q^{NO}$ and $L^{NO}$.
\subsubsection{Basis Functions ILC}
Basis functions ILC achieves reference flexibility by minimizing a cost function and parameterizing the feedforward signal in basis functions as
\begin{equation}
	\label{eq:BFParameterization}
	f_j = \psi\theta_j,
\end{equation}
with feedforward parameters $\theta_j\in\mathbb{R}^{n_\theta \times 1}$ and basis functions $\psi\in\mathbb{R}^{N\times n_\theta}$. For basis functions ILC, the cost function
\begin{equation}
	\label{eq:BFILCCost}
	\begin{aligned}
		\min_{\theta_{j+1}} \left\|\hat{e}_{j+1}\right\|^2_{W_e}+ \left\|f_{j+1}\right\|^2_{W_f}+ \left\|f_{j+1}-f_j\right\|^2_{W_{\Delta f}},
	\end{aligned}
\end{equation}
is minimized. The optimal solution to \eqref{eq:BFILCCost} is of the form
\begin{equation}
	\begin{aligned}
		\theta_{j+1} = Q^{BF}\theta_j+L^{BF}e_j,
	\end{aligned}
\end{equation}
with basis functions ILC robustness and learning matrices $Q^{BF}$ and $L^{BF}$.
\subsubsection{Frequency-Domain ILC}
Frequency-domain ILC iteratively improves the tracking performance by utilizing infinite-time frequency-domain representations. Frequency-domain ILC is designed by the infinite-time update law
\begin{equation}
	\label{eq:fILCphase}
	\mathcal{F}_{j+1}^f(z) = \mathcal{Q}^f(z)(\mathcal{F}_{j}^f(z)+\alpha \mathcal{L}^f(z)\mathcal{E}_j(z)),
\end{equation}
with $\mathcal{Q}^f$ the robustness filter, that is used to enforce convergence and filter out unwanted effects, $\mathcal{L}^f$ the learning filter and $\alpha$ the learning gain. Frequency-domain ILC is implemented in finite-time as \cite{Norrlof2002}
\begin{equation}
	\label{eq:finiteTimefILC}
	f_{j+1}^f = Q^f(f_j^f+\alpha L^f e_j),
\end{equation}
with finite-time convolution matrices $Q^f$ and $L^f$, corresponding to $\mathcal{Q}^f(z)$ and $\mathcal{L}^f(z)$.
\subsection{Problem Definition}
The aforementioned classes of ILC have several design problems and limitations, and are as follows.
\begin{itemize}
	\item For norm-optimal ILC, it is highly complex to robustly choose the weighting matrices, see for example \cite{Xu2002,Gorinevsky2002}, and $W=wI$ severely limits performance. Furthermore, norm-optimal ILC does not have task flexibility.
	\item Basis functions ILC reduces performance if $\psi$ does not accurately describe the inverse system $P^{-1}$ \cite{Boeren2016,VanZundert2016}.
	\item Frequency-domain ILC does have an intuitive design procedure consisting of loop-shaping \cite{Bristow2006,Boeren2016} and results in high performance, but conventionally does not have task flexibility.
\end{itemize}
In \exampleRef{example:compare}, the limited performance of norm-optimal ILC with identity weighting matrices and model uncertainty is illustrated, since robust monotonic convergence is difficult to achieve.
\begin{example}
	\label{example:compare}
	A simulation study illustrates that frequency-domain ILC performs better than norm-optimal ILC for an inaccurate model and identity weighting matrices. The system is a mass-spring-damper as seen in \figRef{fig:twomassystem}, where details are given in \secRef{sec:example}.
	\begin{figure}[tb]
		\centering
		\includegraphics{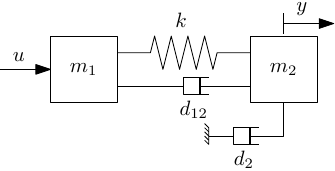}
		\caption{System that is used for validation. The system is discretized with zero-order hold, and has one sample delay.}
		\label{fig:twomassystem}
	\end{figure} %
	The error 2-norm during 300 trials of norm-optimal and frequency-domain ILC is shown in \figRef{fig:comparefILCNOILC1}, and the maximum error 2-norm during these trials and the steady state error 2-norm $\|e_{\infty}\|_2$ are shown in \figRef{fig:comparefILCNOILC}.
	\begin{figure}[tb]
		\centering
		\begin{subfigure}[t]{\linewidth}
			\centering
			\includegraphics{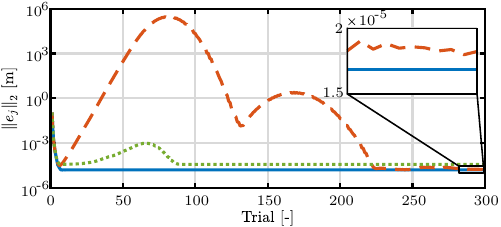}
			\caption{Error 2-norm for frequency-domain ILC \markerline{mblue} and norm-optimal ILC with $W_f=1.39\!\cdot\!10^{-8}I$ \markerline{mgreen}[densely dotted] and $W_f=5.2\!\cdot\! 10^{-9}I$ \markerline{mred}[densely dashed].}
			\label{fig:comparefILCNOILC1}
		\end{subfigure} \hfill \vspace{2mm}
		\begin{subfigure}[t]{\linewidth}
			\centering
			\includegraphics{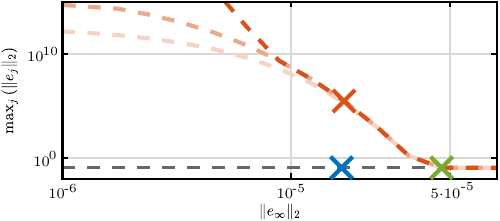}
			\caption{Steady-state and maximum error 2-norm for norm-optimal ILC with $W_f=w_fI \:\forall\: w_f \in [10^{-14},10^{-7}]$ after 75 \markerline{mred}[densely dashed][x][0][1.25pt][8pt][1][0.25], 150 \markerline{mred}[densely dashed][x][0][1.25pt][8pt][1][0.5] and 300 \markerline{mred}[densely dashed] trials, $W_f=5.2\cdot 10^{-9} I$ \markerline{mred}[solid][x][3][1][0] and $W_f=1.9\cdot10^{-8}I$ \markerline{mgreen}[solid][x][3][1][0] correspond to \figRef{fig:comparefILCNOILC1}, frequency-domain ILC \markerline{mblue}[dashed][x][3][1][0] and $f_j=0$ \markerline{gray}[densely dashed][x][0][1.25pt].}
			\label{fig:comparefILCNOILC}
		\end{subfigure}
	\caption{Illustration that norm-optimal ILC with $W_e=I$ and $W_{\Delta f}=0$ converges slowly and non-monotonically for inaccurate models, leading to significantly higher maximum errors $\max_j\left(\|e_j\|_2\right)$.}
	\label{fig:compare}
	\end{figure}
	The results in \figRef{fig:compare} illustrate that for norm-optimal ILC to reduce the steady-state error $\|e_\infty\|_2$ beyond frequency-domain ILC, it first increases the error 2-norm at least a factor $3\cdot10^6$, which is unacceptable in industrial applications.
\end{example}

Hence, the problem addressed in this paper is to develop an ILC algorithm that simultaneously satisfies all three requirements for industrial applicability of ILC, that combines the advantages of current ILC techniques.

\section{Method}
In this section, the developed method is presented. The overparameterized feedforward signal
\begin{equation}
	\label{eq:cilcPara}
	f_j = \Psi \Theta_j = \begin{bmatrix}
		\psi & I_{N}
	\end{bmatrix} \begin{bmatrix}
		\theta_j \\ f^f_j
	\end{bmatrix}=\psi\theta_j+f_j^f,
\end{equation}
that consists of basis functions and frequency-domain ILC, is exploited to achieve both task flexibility and high performance, leading to (R1) and (R2). Frequency-domain ILC is used since it has better performance than norm-optimal ILC for inaccurate models, as shown in \secRef{sec:pdef}, and due to its intuitive design procedure (R3).

First, the norm-optimal description of frequency-domain ILC is determined, such that second, the overparameterized feedforward signal in \eqref{eq:cilcPara} consisting of frequency-domain and basis functions ILC can be jointly optimized. Finally, a procedure summarizes the developed method.
\subsection{Norm-Optimal Representation of Frequency-Domain ILC}
\label{sec:fILCNO}
In this section, it is shown that the finite-time implementation of frequency-domain ILC in \eqref{eq:finiteTimefILC} is equivalent to \eqref{eq:NOILCCost} for a very specific choice of weighting matrices $W_e$, $W_f$ and $W_{\Delta f}$. First, the frequency-domain ILC update is written using the cost function
\begin{equation}
	\label{eq:fILCOptimization}
	\begin{aligned}
		\min_{f_{j+1}^{f}} \left\|\hat{e}_{j+1}\right\|_{W_e^f}^2 \!+\left\|f_{j+1}^{f}\right\|_{W_f^f}^2 \!+\left\|f_{j+1}^{f}-f_j^{f}\right\|_{W_{\Delta f}^f}^2,
	\end{aligned}
\end{equation}
that is minimized by
\begin{equation}
	\label{eq:optimalSolutionfILC}
	\begin{aligned}
		f_{j+1}^f = &(\hat{J}^\top W_e^f \hat{J} +W_f^f+W_{\Delta f}^f)^{-1} \left(\hat{J}^\top W_e^f \hat{J}+W_{\Delta f}^f\right) f_j^f \\
		+&(\hat{J}^\top W_e^f \hat{J} +W_f^f+W_{\Delta f}^f)^{-1}\hat{J}^\top W_e^f e_j \\
		=& Q^f(f_j^f+\alpha L^fe_j),
	\end{aligned}
\end{equation}
where the last identity is found by substituting $f_{j+1}^f$ from the finite-time update law of frequency-domain ILC in \eqref{eq:finiteTimefILC}. The contributions of $f_j^f$ and $e_j$ in \eqref{eq:optimalSolutionfILC} are separated to achieve
\begin{align}
	&Q^f =(\hat{J}^\top W_e^f \hat{J} +W_f^f+W_{\Delta f}^f)^{-1} \left(\hat{J}^\top W_e^f \hat{J}+W_{\Delta f}^f\right), \label{eq:equiv1} \\
	\alpha &Q^fL^f =(\hat{J}^\top W_e^f \hat{J} +W_f^f+W_{\Delta f}^f)^{-1}\hat{J}^\top W_e^f\label{eq:equiv2}.
\end{align}
From \eqref{eq:equiv1}, $W_f^f$ is derived as
\begin{equation}
	\label{eq:Wf1}
	W_f^f = \left(\hat{J}^\top W_e^f \hat{J} + W_{\Delta f}^f\right)\left(\left({Q^f}\right)^{-1}-I\right).
\end{equation}
Because $W_f^f$ must be positive semidefinite, and therefore symmetric, but the product of two symmetric matrices is not necessarily symmetric,  $\left(\hat{J}^\top W_e^f \hat{J} + W_{\Delta f}^f\right)$ in \eqref{eq:Wf1} is chosen
\begin{equation}
	\label{eq:sumIdentity}
	\left(\hat{J}^\top W_e^f \hat{J} + W_{\Delta f}^f\right) = I.
\end{equation}
By substituting \eqref{eq:sumIdentity} and $W_f^f$ from \eqref{eq:Wf1} into \eqref{eq:equiv2}, the following identity is found
\begin{equation}
	\label{eq:finalStepWe}
	\alpha Q^f L^f = Q^f\hat{J}^\top W_e^f,
\end{equation}
leading to the main result in this section in \theoremRef{theorem:weightingMatrices}.
\begin{theorem}
	\label{theorem:weightingMatrices}
	Let $\hat{J}$ be invertible and $L^f=\hat{J}^{-1}$, then the minimizer of the cost function in \eqref{eq:fILCOptimization} with
	\begin{subequations}
	\begin{align}
		\begin{split}
			\label{eq:We}
			&W_e^f = \alpha \hat{J}^{-\top} L^f, 
		\end{split} \\
		\begin{split}
			\label{eq:Wf}
			&W_f^f = \left({Q^f}\right)^{-1}-I,
		\end{split} \\
		\begin{split}
			\label{eq:Wdf}
			&W_{\Delta f}^f = (1-\alpha)I, 
		\end{split}
	\end{align}
\end{subequations}
is equal to finite-time frequency-domain ILC in \eqref{eq:finiteTimefILC}.
\end{theorem}
\begin{IEEEproof}
	Straightforward manipulation of \eqref{eq:finalStepWe} lead to $W_e$ in \eqref{eq:We}. Substituting \eqref{eq:sumIdentity} into \eqref{eq:Wf1} leads to $W_f$ in \eqref{eq:Wf}. Finally, the resulting $W_e$ in \eqref{eq:We} is substituted into \eqref{eq:sumIdentity} to result in $W_{\Delta f}$ in \eqref{eq:Wdf}.
\end{IEEEproof}
\begin{remark}
	\label{rem:2}
	The conditions that $\hat{J}$ is invertible and $L^f=\hat{J}^{-1}$ in \theoremRef{theorem:weightingMatrices} are not restrictive, since the weighting matrix $W_e^f$ can be approximated as the symmetric matrix
	\begin{equation}
		\label{eq:We2}
		\begin{aligned}
			W_e^f = \alpha {L^f}^\top L^f,
		\end{aligned}
	\end{equation}
	since $L^f$ is similar to $J^{-1}$,	and is shown in \secRef{sec:simsetup}.
\end{remark}
\begin{assumption}
	\label{assumption:zerophase}
	To ensure that $W_f^f$ in \eqref{eq:Wf} is symmetric, there is assumed that the robustness filter is designed with zero-phase, i.e., $\mathcal{Q}^f(z)=\mathcal{Q}^f_1(\frac{1}{z})\mathcal{Q}^f_1(z)$ \cite{Gunnarsson2001}.
\end{assumption}
To summarize, finite-time frequency-domain ILC in \eqref{eq:finiteTimefILC} is recovered  by specifically choosing the weighting matrices \eqref{eq:We}, \eqref{eq:Wf} and \eqref{eq:Wdf} and optimizing \eqref{eq:fILCOptimization}, resulting in an intuitive frequency-domain design procedure (R3) for norm-optimal ILC. In the next section, the norm-optimal description of frequency-domain ILC is used when overparameterizing the feedforward signal.
\subsection{Inclusion of Basis-Function in Frequency-Domain ILC}
\label{sec:fILCBF}
In this section, both task flexibility and high performance are achieved by deliberately overparameterizing the feedforward in terms of basis functions and frequency-domain ILC by utilizing the norm-optimal representation of finite-time frequency-domain ILC. The cost function in \eqref{eq:fILCOptimization} is adjusted by utilizing the overparameterized feedforward signal in \eqref{eq:cilcPara} and preserving the weighting $W_f^f$ and $W_{\Delta f}^f$ exclusively on the frequency-domain component as
\begin{equation}
	\label{eq:cILCCostWORegularization}
	\begin{aligned}
		\min_{\Theta_{j+1}} {V}(\Theta_{j+1})  =\min_{\Theta_{j+1}} \left\|\hat{e}_{j+1}\right\|_{W_e^f}^2 +&\left\|\Theta_{j+1}\right\|_{W_{\theta,f}}^2 \\
	+&\left\|\Theta_{j+1}-\Theta_j \right\|_{W_\Delta}^2
	\end{aligned}
\end{equation}
with
\begin{equation}
	\label{eq:wtheta}
	\begin{aligned}
		W_{\theta,f} =  \begin{bmatrix}
			W_{\theta} & 0 \\
			0 & W_f^f
		\end{bmatrix}, && W_{\Delta} = \begin{bmatrix}
		W_{\Delta\theta} & 0 \\
		0 & W_{\Delta f}^f
	\end{bmatrix},
	\end{aligned}
\end{equation}
where $W_{\theta},W_{\Delta\theta} \in\mathbb{R}^{n_\theta\times n_\theta}$  are the weighting matrices on the feedforward parameters $\theta$, that are typically chosen as $W_{\theta} = w_\theta \psi^\top \psi$ and $W_{\Delta \theta} = w_{\Delta \theta} \psi^\top \psi$, resulting in equivalent weighting as conventional basis functions ILC. The minimizer of \eqref{eq:cILCCostWORegularization} is given by
\begin{equation}
	\label{eq:cILCUpdate}
		\begin{aligned}
			\displaystyle
			{\Theta}_{j+1} = &\left(
			\Psi^\top \hat{J}^\top W_e^f\hat{J}\Psi  +W_{\theta,f} + W_{\Delta}
			\right)^{-1} \\
			\cdot &\left(\left(
			\Psi^\top \hat{J}^\top W_e^f \hat{J}\Psi + W_{\Delta}\right) \Theta_j + \Psi^\top \hat{J}^\top W_e^f e_j
			\right).
		\end{aligned}
\end{equation}
\begin{remark}
	\label{rem:complimentary}
	If the weighting on $\theta$ is chosen sufficiently small $W_\theta << W_f^f$, the parameterization \eqref{eq:cilcPara} consisting of frequency-domain and basis functions ILC is naturally complimentary and the parameterization will assign as much information in the basis functions feedforward as possible. Alternatively, a targeted regularization on the frequency-domain component can be done by using the image of $\psi$, similarly to \cite{Kon2023},
	\[
	\min_{\Theta_{j+1}} V(\Theta_{j+1})+\lambda\left\| \begin{bmatrix}
		0 & U_1
	\end{bmatrix} \Theta_{j+1}\right\|_2^2,
	\]
	with singular value decomposition \[\psi=\begin{bmatrix}
		U_1 & U_2
	\end{bmatrix} \begin{bmatrix}
		\Sigma & 0 \\ 0 & 0
	\end{bmatrix}\begin{bmatrix}
		V_1^\top \\ V_2^\top
	\end{bmatrix}.\]
\end{remark}
\subsection{Procedure}
In this section, the developed method for achieving task flexibility, high performance and an intuitive design procedure by combining frequency-domain ILC with basis function ILC is summarized in Procedure~\ref{proc:1}.
\begin{figure}[H]
	\vspace{0mm}
	\hrule 
	\vspace{1mm}
	\begin{proced}[Norm-optimal frequency-domain ILC with basis functions] \hfill \vspace{0.5mm} \hrule \vspace{1mm}
		\label{proc:1}
		\begin{enumerate}
			\item Design learning filter $\mathcal{L}^f(z)$, $\alpha$ and zero-phase robustness filter $\mathcal{Q}^f(z)$ as in frequency-domain ILC.
			\item Derive $L^f$ and $Q^f$, that are the finite-time convolution matrices of $\mathcal{L}^f(z)$ and $\mathcal{Q}^f(z)$.
			\item Choose the basis functions $\psi$ in \eqref{eq:cilcPara}.
			\item Compute equivalent norm-optimal weighting matrices $W_{e}^f$, $W_{f}^f$ and $W_{\Delta f}^f$ using \eqref{eq:We}, \eqref{eq:Wf} and \eqref{eq:Wdf}.
			\begin{enumerate}
				\setlength\itemsep{0.1em}
				\item If $W_e$ is non-symmetrical, follow \remRef{rem:2}.
			\end{enumerate}
			\item Initialize $\Theta_1$, e.g. as $\Theta_1=0$.
			\item For $j\in\{1,2,3,\ldots,N_{trials}\}$.
			\begin{enumerate}
				\setlength\itemsep{0.1em}
				\item Calculate $f_{j}=\Psi\Theta_{j}$ in \eqref{eq:cilcPara}.
				\item Apply $f_j$ to closed-loop system and record $e_j$.
				\item Calculate $\Theta_{j+1}$ using \eqref{eq:cILCUpdate}.
			\end{enumerate}
		\end{enumerate}
		\hrule \vspace{3mm}
	\end{proced}
\end{figure}

\section{Simulation Example}
\label{sec:example}
In this section, the developed method is validated and compared with frequency-domain and basis functions ILC. First, the validation setup is shown, including the system and model, the reference and the ILC designs. Second, the norm-optimal equivalent description of frequency-domain ILC is validated. Finally, the developed method with basis functions is validated and compared for a trial-varying reference.
\subsection{Simulation Setup and Approach}
\label{sec:simsetup}
\begin{figure}[tb]
	\centering
	\includegraphics{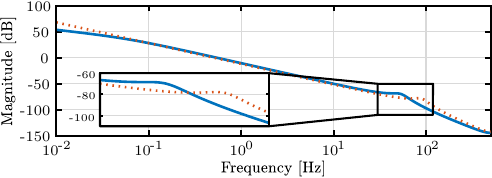}
	\caption{FRF of the system $\mathcal{P}(e^{j\omega})$ \markerline{mblue}[solid] and of the model available for ILC $\widehat{\mathcal{P}}(e^{j\omega})$ \markerline{mred}[dotted].}
	\label{fig:bodes}
\end{figure}

\begin{figure}[tb]
	\centering\includegraphics{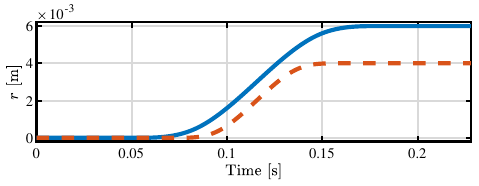}
	\caption{The first \markerline{mblue} and second \markerline{mred}[densely dashed] references that are used during validation.}
	\label{fig:references}
\end{figure}

\begin{figure}[tb]
	\centering
	\includegraphics{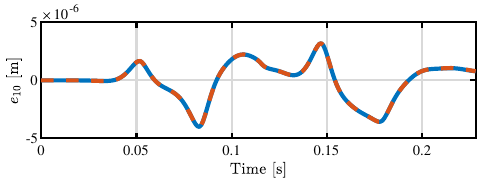}
	\caption{Tracking performance after 10 trials of ILC for frequency-domain ILC \markerline{mblue}[solid] and norm-optimal equivalent \markerline{mred}[dashed] using the first reference, showing the same error profile.}
	\label{fig:difTracking}
\end{figure}

A two-mass-spring-damper system with one sample delay and a sampling time of 1 ms, with inaccurate model, is simulated to validate the developed ILC technique. The system is seen in \figRef{fig:twomassystem}, and the parameters of the true system $\mathcal{P}(z)$ and the model $\widehat{\mathcal{P}}(z)$ are seen in \tabRef{tab:parameters}, and are given by
\begin{equation}
	\resizebox{0.91\linewidth}{!}{$
		\displaystyle
		\begin{aligned}
			\mathcal{P}(z) &=10^{-7}\cdot\frac{2.80z^{-2}+12.4z^{-3}-0.65z^{-4}-1.58z^{-5}}{1 - 3.78 z^{-1} + 5.46 z^{-2} - 3.56 z^{-3} + 0.89 z^{-4}}, \\
			\widehat{\mathcal{P}}(z) &= 10^{-7}\cdot\frac{ 4.00z^{-2} + 21.4z^{-3} + 5.85 z^{-4} - 1.25z^{-5}}{ 1 - 3.56 z^{-1} + 4.98 z^{-2} - 3.26 z^{-3} + 0.85 z^{-4}}.
		\end{aligned}$}
\end{equation}
\begin{table}[tb]
	\centering
	\caption{Parameters used for the system seen in \figRef{fig:twomassystem} for the true system and model.}
	\label{tab:parameters}
	\begin{tabular}{llll}
		\toprule
		\textbf{Parameter} & \textbf{True} & \textbf{Model} & \textbf{Unit}\\
		\midrule
		$m_1$ & 0.072 & 0.09 & [kg] \\
		$m_2$ & 0.01 & 0.006 & [kg]\\
		$k$  & 1000 & 1800 & [N/m] \\
		$d_2$ & 0.031 & 0 & [Ns/m]\\
		$d_{12}$  & 1 & 0.915 & [Ns/m]\\
		\bottomrule
	\end{tabular}
\end{table} %
\hspace{-1mm}The FRF of the systems $\mathcal{P}(e^{j\omega})$ and $\widehat{\mathcal{P}}(e^{j\omega})$ are seen in \figRef{fig:bodes}.
Additionally, the FRF of the true system $\mathcal{P}(e^{j\omega})$ is available for stability analysis, but not for the design of learning filters. The feedback controller is a lead filter and a first-order low-pass filter, that achieves a closed-loop bandwidth of 10 Hz with sufficient robustness margins, and is given by
\begin{equation}
	\begin{aligned}
		\mathcal{K}(z) = \frac{108.6 + 112.9 z^{-1} - 100 z^{-2} - 104.3 z^{-3}}{1 - 0.65 z^{-1} - 0.95 z^{-2} + 0.70 z^{-3}}.
	\end{aligned}
\end{equation}
The learning filter $\mathcal{L}^f(z)$ is designed by approximating the inverse process sensitivity using ZPETC \cite{Tomizuka1987}. The robustness filter $\mathcal{Q}^f(z)$ is a zero-phase second order Butterworth lowpass filter with a cutoff frequency of 40 Hz, that was manually tuned to achieve convergence according to
\begin{equation}
	\label{eq:convfILC}
	\left|\mathcal{Q}^f(e^{j\omega})(1-\alpha\mathcal{J}(e^{j\omega}) \mathcal{L}^f(e^{j\omega}))\right|<1, \quad \forall \omega \in[0,2 \pi].
\end{equation}
Two different references are used to show task flexibility, where both references are performed for 10 consecutive trials. The references are fourth-order polynomial references designed using the approach in \cite{Lambrechts2005} with $N=229$ samples and are seen in \figRef{fig:references}.
\paragraph{Basis Function Design}
The basis functions $\psi$ are based on the inverse model of $\widehat{\mathcal{P}}(z)$ and are chosen as
\begin{equation}
	\label{eq:psichoice}
	\psi = \begin{bmatrix}
		\ddot{r} & \dddot{r} & \ddddot{r}
	\end{bmatrix},
\end{equation}
where the derivatives of the reference are readily available by design of the reference trajectory. $W_\theta$ and $W_{\Delta \theta}$ in \eqref{eq:wtheta} are chosen as $0$, since no additional robustness is necessary.
\paragraph{Recovering Norm-Optimal Formulation of Frequency-Domain ILC} $W_f^f$ and $W_{\Delta f}^f$ are calculated using respectively \eqref{eq:Wf} and \eqref{eq:Wdf}, resulting in $W_{\Delta f}^f=0$ since $\alpha=1$. $W_e^f$ is computed using \eqref{eq:We2}, since $L^f\neq \hat{J}^{-1}$ due to the use of ZPETC, as indicated in \remRef{rem:2}. Frequency-domain ILC and its norm-optimal equivalent achieve the same tracking error after 10 trials as seen in \figRef{fig:difTracking}, validating their equivalence.
\subsection{Validation Results}
\begin{figure}[tb]
	\centering
	\includegraphics{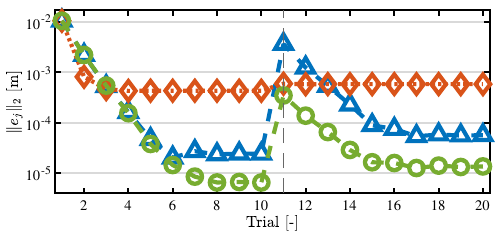}
	\caption{Error 2-norm for 20 trials of frequency-domain ILC \markerline{mblue}[dashed][triangle], basis function ILC \markerline{mred}[dotted][diamond] and developed combined frequency-domain and basis function ILC \markerline{mgreen}[dashed][o]. At trial 11, the reference is changed \markerline{black}[dashed][][2][0.75].}
	\label{fig:2normResult}
\end{figure}

\begin{figure}[tb]
	\centering
	\includegraphics{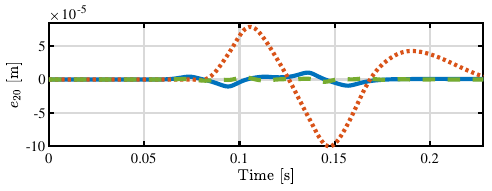}
	\caption{Tracking error after 20 trials $e_{20}$ of frequency-domain ILC \markerline{mblue}[solid], basis function ILC \markerline{mred}[dotted] and developed combined frequency-domain and basis function ILC \markerline{mgreen}[dashed].}
	\label{fig:timeError}
\end{figure}

\begin{figure}[tb]
	\centering
\includegraphics[width=\linewidth]{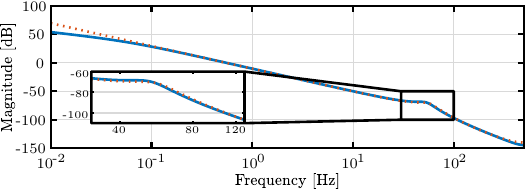}
	\caption{FRFs of system $\mathcal{P}(e^{j\omega})$ \markerline{mblue} and of estimate using the inverse basis function feedforward model $\mathcal{F}_{BF}^{-1}(e^{j\omega},\theta_j)$ \markerline{mred}[dotted], with finite-time representation ${F}_{BF}(\theta_j)r = \psi\theta_j$.}
	\label{fig:bodeFit}
\end{figure}

The error for 20 trials of ILC is seen in \figRef{fig:2normResult} and \figRef{fig:timeError}. The plant estimate using the basis function feedforward is seen in \figRef{fig:bodeFit}.
The following observations are made.
\begin{itemize}
	\item \figRef{fig:2normResult} shows frequency-domain ILC converges at trials 10 and 20, but its error increases at trial 11, showing that it lacks task flexibility.
	\item Though basis functions ILC enables reference flexibility as seen in \figRef{fig:2normResult}, its higher error compared to the other methods stems from lacking $\dot{r}$ in its basis function $\psi$ to compensate the viscous friction $d_2$.
	\item The error at trial 20 in \figRef{fig:2normResult} and \figRef{fig:timeError} demonstrate the developed approach's superior performance against both basis function and frequency-domain ILC. It surpasses basis functions by handling friction with the frequency-domain component, and outperforms frequency-domain ILC by capturing high-frequency effects with $\psi$, that for frequency-domain ILC is filtered out by $\mathcal{Q}^f(z)$.
	\item Similar error 2-norm to basis functions ILC under reference change illustrates the method's task flexibility.
	\item From \figRef{fig:bodeFit} it becomes clear that the basis functions feedforward parameters $\theta_j$ are estimated consistently with the inverse model, which is enabled since $W_\theta=0<<W_f^f$ as described in \remRef{rem:complimentary}.
\end{itemize}
\section{Conclusions}
In this paper, both task flexibility and performance is achieved through the use of an overparameterized feedforward signal consisting of frequency-domain and basis functions ILC. The finite-time norm-optimal representation of frequency-domain ILC is derived, that is consequently used in overparameterizing the feedforward signal. The basis functions and freqency-domain ILC components are complimentary by appropriately regularizing the frequency-domain component. An example validates the equivalent norm-optimal representation, and by exploiting the overparameterized feedforward signal, the performance is significantly increased. Hence, the developed method is a key enabler for improving performance and task flexibility in control.	

Ongoing research is aimed at verifying the developed method on an experimental setup and computationally efficient implementations.
\bibliographystyle{IEEEtran}
\bibliography{../../library,BSTControl}


\end{document}